\documentclass{epl}

\title{Influence of the cosmological expansion on small systems}
\shorttitle{Small systems and the cosmological expansion}

\author{A. Dom{\'\i}nguez\inst{1}\thanks{E-mail:
    \email{alvaro@theorie.physik.uni-muenchen.de}} \and J.
  Gaite\inst{2}\thanks{E-mail: \email{gaite@laeff.esa.es}}}
\institute{   
  \inst{1} Theoretische Physik, Ludwig-Maximilians-Universit\"at --
  Theresienstr.~37, D-80333 M\"unchen, Germany \\
  \inst{2} Centro de Astrobiolog{\'\i}a, Instituto Nacional de
  T{\'e}cnica Aeroespacial -- Ctra.~de To\-rre\-j{\'o}n a Ajalvir, 28850
  Torrej{\'o}n de Ardoz, Madrid, Spain
  } 
\shortauthor{A. Dom{\'\i}nguez \etal}


\begin{document}

\maketitle

\begin{abstract}
  The effect of the large-scale cosmological expansion on small
  systems is studied in the light of modern cosmological models of
  large-scale structure. We identify certain assumptions of earlier
  works which render them unrealistic regarding these cosmological
  models.  The question is reanalyzed by dropping these assumptions to
  conclude that a given small system can experience either an
  expansion or a contraction of cosmological origin.
\end{abstract}

Since Hubble proposed his interpretation of the cosmological redshift
as expansion of the Universe, there is an ongoing debate about what
scales actually participate in this expansion, a problem which has
been reconsidered by different authors since then (see
Refs.~\cite{EiSt,DiPe,NoPe,Ande,CFV,Bonn,SMV}, and further references
therein). We will address the problem by relaxing some previous
assumptions to get a model in better agreement with the current
cosmological picture. We then find that the large-scale expansion does
imply a (minute) time dependence of the volume of small systems, but
also that this change can be, contrary to previous analysis, either an
expansion or a contraction: the reason is that the origin of this
effect is conceptually different from the origin of the effect studied
in previous works.
  
Refs.~\cite{DiPe,NoPe,Ande,CFV,Bonn} consider the simplified model of
a small bound system embedded on a Robertson-Walker (RW) metric: we
call this ``the RW picture''. By ``small'' is meant a size much
shorter than the scales characteristic of the background RW metric,
i.e., the horizon and the radius of curvature (e.g., it is customary
to consider an atom, a planetary system or a galaxy). The effect of
the background RW metric is then considered a small perturbation on
the dynamical evolution of the small system, treated itself as a {\em
  test particle} independent of the RW background:
Refs.~\cite{DiPe,Ande} state that a small bound system suffers no
expansion, because they simply consider the zeroth-order, unperturbed
evolution; Refs.~\cite{NoPe,CFV,Bonn} compute the first correction,
which yields an expansion of the system (albeit much slower than the
cosmological expansion).  However, the RW picture involves two
important assumptions contrary to modern cosmological models: (i) via
the Einstein equations, there must exist some sort of stress-energy
source that can be considered uniformly distributed and Hubble-flowing
even at the small scale of the bound system, (ii) the gravity pull
exerted by this source must dominate over the self-gravity of the
system, i.e., the effect of the system's stress-energy on the RW
metric is a small perturbation. This latter assumption excludes the
case in which the small system is bound by its own gravity, but it can
be relaxed \cite{NoPe}, as we shall describe below.
  
Let us examine if these assumptions are upheld by the prevailing
cosmological models \cite{cosmo}. These models predict a ``bottom-up''
scenario of structure formation: due to the gravitational instability,
the matter driving the cosmological expansion on the large scales
ceases to Hubble-flow (it ``decouples'' from the expansion) and
collapses, becoming inhomogeneously distributed below some scale $R$,
which grows with time and today is certainly larger than the galactic
scales. Thus the Hubble flow just plays the role of an initial
condition which is gradually destroyed on ever larger scales by the
gravitational instability. This result contradicts assumption (i).
However, one cannot exclude the presence of a uniform and still
Hubble-flowing background of relativistic particles (in
Ref.~\cite{NoPe} were mentioned photons, very light neutrinos and
gravitational waves, but there can be others).  Nevertheless, this
background source, with density parameter $\Omega \ll 1$, does not
fulfill the condition (ii): the source of self-gravity corresponds to
a density parameter that ranges from $\Omega \sim 10^{3}-10^{5}$ for a
galaxy to $\Omega \sim 10^{30}$ for a hydrogen atom.
  
This discussion clearly shows that the RW picture is only of academic
interest and has little use for a realistic study of a possible
small-scale expansion: the metric on small scales does not look at all
like the large-scale RW metric, but rather corresponds to that of an
inhomogeneous distribution of bound systems. In fact, the interest is
shifting precisely towards an appreciation of this small-scale
graininess: Whereas the old question was ``how can there be
constant-volume systems in an expanding Universe?'', the new question
is rather the reverse ``how can there be a large-scale RW metric at
all given the pervading small-scale graininess?''. In other words,
since the Einstein equations are nonlinear, it is not granted that
they preserve their form upon coarse-graining and it can therefore be
questioned whether the RW metric really represents the large-scale
solution despite that the matter distribution looks homogeneous when
smoothed on large scales \cite{aver}. In any event, we will not
address this new question, and we confine ourselves to answer the old
question with due consideration for the small-scale graininess.

In connection with these arguments, the results following from the RW
picture can possibly be reinterpreted as an average: on the small
scales, the cosmological expansion is only an average property of a
grainy Universe and it could be perceived by the small system only by
placing it at every location in the Universe. However, such
interpretation is also unrealistic because the small system cannot be
treated as a simple test particle, that can be anywhere: the small
systems of interest belong always to high density collapsed regions
(clusters, galaxies, gas clouds, \dots). Thus, the averaging procedure
should restrict the small system to being placed in these regions,
which, according to the modern scenarios of structure formation, no
longer expand. That is, the effect of the {\em local} environment on
the small system will be unrelated to the universal expansion and the
system will perceive this expansion, as Hubble himself did, only
through its interaction with the receding distant matter, i.e.,
farther than the scale $R$ mentioned above.  Therefore, we consider
the following simplified model, which, in view of the previous
discussion, is more suitable than the RW picture: We begin with a set
of pointlike masses (galaxies, say) homogeneously distributed and
Hubble-flowing and make a spherical cavity of fixed radius $R$
centered on one of these pointlike masses which is assumed to contain
the bound system.  Refs.~\cite{EiSt,DiPe} consider this kind of model
to conclude that the small system does not suffer any kind of
expansion because the distribution of inhomogeneities outside the
cavity has an effective spherical symmetry. However, {\em this}
symmetry holds only on average, so that in principle there could be a
nonvanishing effect on the small system due to the deviation from
perfect isotropy. Our goal, therefore, is to study the influence of
the (Hubble-flowing) anisotropies on a small bound system and, in
particular, whether they give rise to a change of its volume.
 
Let $L$ denote the size of the small system and $d_\ab{H}$ the Hubble
radius of the large-scale effective RW metric associated with the
Hubble-flowing pointlike masses (galaxies). One then has the
inequalities $L \ll R \ll d_\ab{H}$, representing that the system is small
and that the Hubble flow is observable.  The assumption of pointlike
masses implicitly requires that their size is negligible compared to
$R$. At this point, we introduce an important simplification: rather
than carrying out a fully relativistic analysis, we will work in the
Newtonian limit in the reference frame of the small bound system. This
is justified by the above mentioned condition that $d_\ab{H}$ is much
larger than the physically interesting length scales and by the
assumptions that the involved velocities are non-relativistic and that
there are no singularities in the metric. In fact, we are working
within the same approximation customarily employed to study the
formation of large-scale structure in cosmological models
\cite{cosmo}.
  
The fundamental quantity now is the gravitational potential $\phi
({\bm r}, t)$ created by the distant inhomogeneities. By virtue of the
superposition principle of Newtonian gravity, we can add a uniform,
Hubble-flowing matter distribution as another source of $\phi$: this
will allow us to consider also the RW picture, for comparison with
previous works. The Newtonian equation of motion of the particles
forming the small bound system is simply (${\bm r}$ is the position of
any one particle)
\begin{equation}
  \label{newton}
  m \frac{\upd^2 r_i}{\upd t^2} = F_i ({\bm r}) - 
  m \phi_{,i} ({\bm r}, t) ,
\end{equation}
where $F_i$ collects all the internal forces acting between particles
within the system (including their mutual gravitational attraction),
which keeps the system bound. The effect of the potential is
considered as a perturbation over the evolution driven by the internal
forces. Since the potential varies over length scales much larger than
the size $L$ of the system, one can Taylor-expand $\phi ({\bm r})$
around the coordinate origin ${\bm r}={\bm 0}$. The homogeneous
gravitational acceleration term $\phi_{,i} ({\bm 0})$ can be
eliminated by taking the coordinate origin at the position of the
free-falling center of mass of the system: then, the lowest order term
in this expansion corresponds to the tidal stresses, $r_j \phi_{,ij}
({\bm 0})$ (hereafter, we assume a summation over pairwise repeated
indices) \cite{f1}.  Furthermore, we can also assume that the
characteristic time scales of the small system are much shorter than
the cosmological scales of temporal variation of $\phi$. This allows
an adiabatic approximation, by which we can compute the response of
the system as if the gravitational field were static and afterwards
introduce its time dependence. Proceeding further requires a
specification of the internal forces. We consider a many-particle
system that can be modelled as a continuum and so Eq.~(\ref{newton})
must be appropriately rewritten. This continuum model is an innovation
over previous works \cite{EiSt,DiPe,NoPe,Ande,CFV,Bonn,SMV}.  While it
somewhat simplifies the reasoning, it keeps the essential features,
and the conclusions can be easily extrapolated to systems such as an
atom or a planetary system.
  
We then consider the differential equation expressing equilibrium
under the {\em small} deformations induced by the tidal forces
\cite{LaLiElast},
\begin{equation}
  \label{equil}
  \sigma_{ij,j} ({\bm r}) = \varrho_0 ({\bm r}) r_j 
  \phi_{,ij} ({\bm 0}) , 
\end{equation}
where $\sigma_{ij} ({\bm r})$ is the stress tensor field induced by
the deformation and $\varrho_0 ({\bm r})$ is the mass density field of
the unperturbed body (density perturbations yield a term of second
order in the perturbing tidal forces in Eq.~(\ref{equil})).  Given
that the deformation must be small, $\sigma_{ij}$ will be a linear,
local function of the displacement field $u_i ({\bm r})$, so that this
equation reduces to an inhomogenous, linear equation for this field. A
great simplification of this equation is achieved by considering an
elastic, isotropic solid, so that Hooke's law applies
\cite{LaLiElast}:
\begin{equation}
  \label{hooke}
  \sigma_{ij} = \left(K - \frac{2}{3} \mu \right) u_{k,k} \delta_{ij} 
  + \mu (u_{i,j} + u_{j,i}) ,
\end{equation}
where $K$ and $\mu$ are the elastic constants of the solid. Besides,
we are only interested in the change of volume of the system due to
the perturbing potential. Let $V_0$ denote the volume of the
undeformed body, and $V$ that after deformation. The volume change
$\delta V \equiv V-V_0$ is the integral over the body of the field
$u_{k,k}({\bm r})$. This latter field can be shown to obey a Poisson
Eq.\ by taking the $i$-derivative of Eq.~(\ref{equil}) and using
Eq.~(\ref{hooke}). This equation is easily solved with the natural
boundary condition $u_k \equiv 0$ if $\phi \equiv 0$. One then finds
$\delta V = C \tens{D}_{ij} \phi_{,ij} ({\bm 0})$, with the
proportionality factor $C=1/(3 K + 4 \mu) >0$ (basically the
isothermal compressibility) and the second-rank tensor
\begin{equation}
  \label{tensD}
  \tens{D}_{ij} = - \frac{3}{4 \pi} \int_{V_0} \upd{\bm r} \upd{\bm r}' \, 
  \frac{(\varrho_0 ({\bm r}) r_j)_{,i}}{|{\bm r}-{\bm r}'|} = 
  - \frac{3}{4 \pi} \int_{V_0} \upd{\bm r} \, 
  \varrho_0 ({\bm r}) r_j \int_{V_0} \upd{\bm r}' \frac{\partial}{\partial r'_i} 
  \frac{1}{|{\bm r}-{\bm r}'|}.
\end{equation}
We further make the approximation of taking the body quasi-spherical
(e.g., a star or a planet) to achieve mathematically simple
expressions without sacrificing the physical content of the problem we
are discussing. The ${\bm r}'$-integral is trivial for a spherical
$V_0$, and one can show after some algebra that
\begin{equation}
  \label{deltaV}
  \delta V = C \tens{I}_{ij} [\phi_{,ij} ({\bm 0}) - 
  \frac{1}{3} \phi_{,kk} ({\bm 0}) \delta_{ij}] - 
  \frac{1}{6} C \tens{I}_{ii} \phi_{,jj} ({\bm 0}) , 
\end{equation}
where 
\begin{equation}
  \tens{I}_{ij} = \int_{V_0} \upd{\bm r} \, (r^2 \delta_{ij} - r_i r_j) 
  \varrho_0 ({\bm r})
\end{equation}
is the moment of inertia tensor of the undeformed body \cite{Gold}.
  
We have written the change of volume as the sum of the changes induced
separately by the traceless tidal tensor and by its trace,
respectively. The latter, because of the Poisson equation, is
determined by the matter distribution at the position of the system
(i.e., the smoothly distributed component in our case) and represents
always a decrease of volume: the mass of the uniform component
enclosed within the system causes an inward attraction which contracts
the body with respect to its unperturbed state. The traceless tidal
tensor, however, carries information about the distant matter
distribution and does not produce a change of definite sign: it can
represent either a contraction or an expansion; this will depend on
the orientation of the body relative to the pointlike sources.  Notice
that this latter change does not arise either if the traceless tidal
tensor vanishes (e.g., in the case of a perfectly isotropic matter
distribution, as in the RW picture) or if the unperturbed body has
{\em exact} spherical symmetry ($\tens{I}_{ij} \propto \delta_{ij}$).
  
The effect considered is static and no mention of an expanding
Universe has been made so far. This enters through the time dependence
of the tidal forces. The adiabaticity assumption implies that this
temporal dependence is so slow that equilibrium, Eq.~(\ref{equil}),
holds at any instant. We can then just take expression (\ref{deltaV})
and include the time dependence of $\phi_{,ij}({\bm 0})$.  A local
expansion, if any, is measured by the time derivative of $\delta V$,
thus comparing the volume of the body at successive instants, rather
than with the volume of the unperturbed body. In the simple models we
consider below, it is $\phi_{,ij} ({\bm 0}, t) \propto a(t)^{-3}$,
where $a(t)$ denotes the expansion factor, due to the dilution of the
total (uniform + pointlike components) mass density by the Hubble
flow. It then follows from Eq.~(\ref{deltaV}) that $(\delta V)\dot{} =
- 3 H \delta V$, where the dot denotes temporal derivative and
$H=\dot{a}/a$ is the Hubble function. Following \cite{NoPe}, we can
define a local, effective Hubble constant, $\tilde{H}$, as
\begin{equation}
  \label{Heff}
  \tilde{H} \equiv \frac{\dot{V}}{3 V} = 
  \frac{(\delta V)\dot{}}{3 (V_0 + \delta V)} 
  \approx - \frac{\delta V}{V_0} H .
\end{equation}
This is the main result of the analysis: the effective Hubble constant
is much smaller than the true Hubble constant in absolute value
(because the deformation is small, $|\delta V| \ll V_0$), and it need
not even be positive, indicating that the expanding Universe could
give rise to a local contraction.
  
We now consider a couple of applications of the result~(\ref{Heff}).
First, to compare with the RW picture, we disregard inhomogeneities
altogether and take into account only the effect of a uniform
component with mass density $\varrho_b (t) \propto a(t)^{-3}$. Then,
one has $\phi ({\bm r}, t) = (2 \pi/3) G \varrho_b (t) r^2$ and the
tidal tensor $\phi_{,ij} = (4 \pi G \varrho_b/3) \delta_{ij}$ is a
completely local quantity. Hence \cite{f2}
\begin{equation}
  \label{Hrw}
  \tilde{H}_\ab{RW} = \frac{2 \pi C G \tens{I}_{ii} \varrho_b}{3 V_0} H > 0.
\end{equation}
Therefore, one has a local expansion, albeit much slower than the
Hubble flow. For example, taking $\varrho_b \sim 10^{-26}\un{kg\,
  m^{-3}}$ (corresponding to $\Omega \sim 1$) and values appropriate
for the planet Earth ($M \sim 10^{24}\un{kg}$, $R \sim 10^7\un{m}$,
and $C \sim 10^{-11}\un{N^{-1}m^2}$ suitable for typical solids
\cite{Kitt}), one finds $\tilde{H}_\ab{RW} \sim 10^{-30} H$. This
conclusion agrees qualitatively with that obtained in
Refs.~\cite{NoPe,Bonn,CFV} for a system consisting of two particles in
a Keplerian orbit (the detailed numerical values depend on the
particular system, i.e., on the typical value of the internal forces
$F_i$ in Eq.~(\ref{newton})). The result~(\ref{Hrw}) could be
reinterpreted considering $\varrho_b$ to be the spatial average of the
actual inhomogeneous density. But, as emphasized, the RW picture is
inconsistent with present cosmological models. To address the effect
of the cosmological expansion, one should rather use the more general
result following from Eq.~(\ref{deltaV}) and consider instead the
effect of the traceless tidal tensor.
  
We study, therefore, the alternative model introduced before: there is
no uniform component at all but only Hubble-flowing, pointlike masses
beyond some {\em fixed} distance $R$ from the small system. Thus, the
density field observed by the small system can be written as $\varrho
({\bm r}, t) = \sum_\alpha M \delta [{\bm r}- a(t) {\bm
  r}^{(\alpha)}]$, with $|{\bm r}^{(\alpha)}| > R$. Here, $M$, ${\bm
  r}^{(\alpha)}$ refer to the mass and position at the present time of
each pointlike source, respectively, while the scale factor $a(t)$
(with $a(\ab{now})=1$) accounts for the Hubble law. This density field
then yields
\begin{equation}
  \label{farphi}
  \phi_{,ij} ({\bm 0}, t) = \frac{G}{a(t)^3} \int_{x>R} \!\!\!\! 
  \upd{\bm x} \, \frac{\varrho({\bm x}, t_\ab{now})}{|{\bm x}|^3} 
  \left\{ \delta_{ij} - \frac{3 x_i x_j}{|{\bm x}|^2} \right\} .
\end{equation}
Since $\phi_{,ij}$ is now traceless, taking as coordinate axis the
principal axis of the tensor $\tens{I}_{ij}$ and denoting by $I_{i}$ the
principal moments of inertia, Eqs.~(\ref{deltaV}) and (\ref{Heff})
yield
\begin{equation}
  \label{Heff2}
  \tilde{H} = - \frac{C \sum_i I_i \phi_{,ii} ({\bm 0})}{V_0} H .
\end{equation}
This effective Hubble constant depends on the specific positions ${\bm
  r}^{(\alpha)}$ of the sources; an estimate of its typical value can
be gained by averaging over different realizations of the distribution
of sources {\em with the constraint that there is one at ${\bm
    x}={\bm0}$}. Let $\langle \cdots \rangle_c$ denote this
conditional average. We then have, assuming statistical homogeneity
and isotropy, \cite{cosmo}
\begin{eqnarray}
  \label{correl}
  \langle \varrho ({\bm x}) \rangle_c = \varrho_b [1 + \xi (|{\bm x}|)] , \\
  \langle \varrho ({\bm x}) \varrho ({\bm y}) \rangle_c = \varrho_b^2 
  [1 + \xi (|{\bm x}|) + \xi (|{\bm y}|) + \xi (|{\bm x}-{\bm y}|) + 
  \zeta (|{\bm x}|, |{\bm y}|, |{\bm x}-{\bm y}|)] , \nonumber
\end{eqnarray}
where we have identified the (unconditional) average with $\varrho_b$,
the mean density driving the large-scale expansion, and $\xi$ and
$\zeta$ are, respectively, the two and three-point reduced correlation
functions \cite{f3}. Expressions (\ref{farphi}) and (\ref{Heff2}) now
yield $\langle \tilde{H} \rangle_c = 0$. This result is not
surprising: because of statistical isotropy in the distribution of
distant matter, there is no effect at all on average, and so we
recover the result in Refs.~\cite{EiSt,DiPe}.  However, for a {\em
  given} system at a {\em given} position, there will be a
non-vanishing $\tilde{H}$, of which the order of magnitude can be
estimated from its variance, $\langle \tilde{H}^2 \rangle_c$.  We
define the ratio $\eta^2 = \langle \tilde{H}^2
\rangle_c/(\tilde{H}_\ab{RW})^2$ as a measure of this effect relative
to the prediction of the RW picture, Eq.~(\ref{Hrw}). The exact
expression for $\eta$ follows from Eqs.~(\ref{Hrw}-\ref{correl}):
\begin{equation}
  \label{eta}
  \eta^2 = \frac{9}{(2 \pi)^2} \int_{x, y>R} \!\!\!\! \upd{\bm x} \upd{\bm y} \, 
  \frac{\xi (|{\bm x}-{\bm y}|)+\zeta (|{\bm x}|, |{\bm y}|, |{\bm x}-{\bm y}|)}
  {|{\bm x}|^5 |{\bm y}|^5} (x_i^2 \lambda_i)(y_j^2 \lambda_j) ,
\end{equation}
where $\lambda_i= (3 I_i / \sum_k I_k) - 1$ is a measure of the
departure of the unperturbed body's shape from a sphere.  To estimate
this ratio for a given body, we need a specific form of $\xi$ and
$\zeta$.  We can assume that galaxies are not-too-biased tracers of
the total mass distribution and have approximately the same mass. The
two-point galaxy correlation extracted from catalogs can be
approximated by $\xi (r) \approx (r_0/r)^\gamma$, $\gamma \approx 1.7$
over a wide range of scales, from about $50\un{kpc}$ up to
$20\un{Mpc}$ \cite{cosmo}. The scale $r_0$ physically represents the
smoothing scale above which the smoothed density field of the
pointlike sources appears homogeneous.  In the standard cold dark
matter (CDM) models, it is this large-scale homogeneous density which
rules the cosmological expansion, and so one expects that the Hubble
law holds also above this scale. And indeed, the standard conclusion
from observations is that both $r_0$, $R$ are of the order of the
megaparsec \cite{cosmo}. Moreover, on large scales, $r>r_0 \approx R$,
the correlations are small and asymptotically $1 \gg |\xi| \gg
|\zeta|$, so it can be argued that the contribution of $\zeta$ to the
integral~(\ref{eta}) is at most of the same order as that of $\xi$.
Therefore, under the assumption that the main contribution to the
integral in Eq.~(\ref{eta}) arises from the range of scales where
$\xi$ follows the power law, one can easily estimate the order of
magnitude of $\eta$,
\begin{equation}
  \label{ratio}
  \eta^2 \sim \lambda^2 \left( \frac{r_0}{R} \right)^\gamma 
  \int_{x, y>1}   \frac{\upd{\bm x} \upd{\bm y}}
{|{\bm x}|^3 |{\bm y}|^3 |{\bm x}-{\bm y}|^\gamma} ,
\end{equation}
where $|\lambda|$ denotes the order of magnitude of $\lambda_i$, and
the integral is convergent for $0<\gamma<3$ and of the order of unity.
Consequently, if this result were extrapolated to a not-too-spherical
body (with $|\lambda| \sim 1$, e.g., a rod or a disc), the local
Hubble constant due to the distant sources would be of the same order
as the prediction of the RW picture; the important difference is that
$\tilde{H}$ is a fluctuating quantity that can take either sign. On
the other side, for a (quasi-spherical) planet the local Hubble
constant is considerably reduced. For example, regarding the Earth
again, it is reduced by a factor $|\lambda| \sim 10^{-3}$ \cite{Gold}.
 
It has been recently claimed \cite{SMP} that the power-law behavior
holds over the whole range of scales probed by galaxy catalogs, and
$r_0$ ($> 100\un{h^{-1}Mpc}$) cannot be determined from the avalaible
catalogs yet, so that we face what seems a fractal Universe. These
authors, however, concede that the Hubble law already holds for $R$ of
the order of the megaparsec, so that $r_0 \gg R$, which constitutes
what they call the {\em Hubble-de Vaucoleurs paradox}. If this
interpretation proves right, then our previous reasoning would yield a
value of $\eta/\lambda$ much larger than unity (and dominated by the
strong three-point correlation), so the local Hubble constant could
therefore be much larger than the prediction according to the standard
CDM models.

In conclusion, we have argued that the customary RW picture is not
suitable to address the possible influence of the cosmological
expansion on small systems. We have proposed a conceptually new,
improved picture which takes into account the observed small-scale
inhomogeneity in the matter distribution and the fact that the {\em
  relevant} small systems cannot be considered simple test particles,
since they are themselves part of the Universe and belong to collapsed
structures. This has enabled us to analyze this effect within the
framework of realistic cosmological models. The study has been carried
out in the Newtonian limit of General Relativity, which already
encompasses the correct physics of the problem and provides a clear
and simple explanation of how the cosmological expansion may affect
small systems. We deduce that no small-scale expansion occurs {\em on
  average} due to statistical isotropy of the matter distribution, but
at a given position in space, tidal effects due to the distant,
Hubble-flowing inhomogeneities may give rise either to an {\em
  expansion} or a {\em contraction} in time of the volume of a small
system. This effect is, as expected, of a very small magnitude (e.g.,
for the Earth about $10^{33}$ times smaller than the cosmological
expansion), so that, for all practical purposes, it can be safely
ignored.

\acknowledgments 

We acknowledge helpful discussions with J.\ M.  Mart\'\i n--Garc\'\i
a. We also thank the Laboratory for Spatial Astrophysics and
Fundamental Physics (LAEFF, Spain), where most of the research
conducive to this paper was done.


\begin{thebibliography}{99}
\bibitem{f1} When $\phi_{ij} ({\bm 0})$ is computed for the RW picture
  (see right before Eq.~(\ref{Hrw})), Eq.~(\ref{newton}) then becomes
  the Newtonian limit of the equation for geodesic deviation in Fermi
  normal coordinates, as computed in Ref.~\cite{CFV}.
\bibitem{f2} The presence of a cosmological constant would contribute
  in Eq.~(\ref{deltaV}). But, for a time-independent cosmological
  constant, this contribution is just a static effect and does not
  affect the effective Hubble constant, Eq.~(\ref{Heff}).
\bibitem{f3} In Eqs.~(\ref{correl}), we have neglected shot-noise
  contributions from the pointlike nature of the density distribution.
  This is justified when the average number of points in the volumes
  of interest (in our case, of order $R^3$) is very large.
\end{thebibliography}
\end{document}